\documentstyle[11pt,newpasp,twoside]{article}
\def\edcomment#1{\iffalse\marginpar{\raggedright\sl#1\/}\else\relax\fi}
\reversemarginpar

\begin{document}
\title{Low-ionization structures in planetary nebulae}
 \author{D. R. Gon\c calves, R. M. L. Corradi, E. Villaver, A. Mampaso}
\affil{Instituto de Astrof\'\i sica de Canarias, E-38200 La Laguna, Tenerife, 
Spain}
\author{and M. Perinotto}
\affil{Dipartamento di Astrof\'\i sica e Scienza dello Spazio, Universit\`a 
di Firenze, Largo E. Fermi 5, I-50125 Firenze, Italy}

\begin{abstract}
We present new results of a program aimed at studying the physical properties,
origin and evolution of those phenomena which go under the somewhat generic
definition of ``low-ionization, small-scale structures in PNe". We have obtained
morphological and kinematical data for 10 PNe, finding low-ionization 
structures
with very different properties relative to each other,  in terms of
expansion velocities, shapes, sizes and 
locations
relatively to the main nebular components.  It is clear that several physical
processes have to be considered in order to account for the formation and
evolution of the different structures observed. 
We present here some results that are illustrative of
 our work -- on IC 4593, NGC 3918, K 1-2, Wray 17-1, 
NGC 6337, He 2-186 and K 4-47 -- and some of the questions that we try 
to address.
\end{abstract}

\section{Introduction}
The presence of low-ionization structures (LIS) in PNe is poorly understood. 
They 
appear as jets, knots, filaments and tails, attached to or detached from
 the main 
shell.  They are often labeled with specific acronyms trying to describe 
some of their characteristics -- for instance, FLIERs: fast, low-ionization 
emission regions (Balick et al. 1993); or  
BRETs: bipolar, rotating, episodic jets (L\'opez, V\'azquez, \& Rodr\'\i guez 
 1995). Despite  these pieces of information, the only clear characteristic  
of all PNe containing LIS is the presence of features much more prominent 
in low-ionization lines (such as [NII] and [SII]) than in the more highly
ionized ones 
(typically H${\alpha}$ and [OIII]).

There exist some 
families of models trying to explain the origin of the LIS. Basically, 
these are: interaction stellar wind models (Frank, Balick, \& Livio 1996; 
 G\'arcia-Segura et al. 1999); 
jet interaction with the circumstellar medium (Cliffe et al. 1995), and the 
interaction of the shell with the interstellar medium  
(Soker \& Zucker 1997). See Mellema (1996) for a review of LIS and 
their models. In addition, some ingredients -- such as stellar magnetic 
fields, rotation; precession, a binary system in the center, and 
dynamical (Kelvin-Helmholtz and Rayleigh-Taylor) and/or radiation 
in situ instabilities -- can be considered in these families of models in 
order to try to match the observations. 

\section{Results for two samples of PNe with LIS}

\subsection{Literature sample}

We searched in the literature for PNe which present LIS, finding about 50 
objects. From this sample $\sim 50\%$ have jets or jet-like structures and 
$\sim 35\%$ 
present any other kind of symmetric LIS (radially symmetrical or 
point-symmetric pairs). The remaining $\sim 15\%$ are PNe 
that show LIS more or less oriented in the radial direction, probably being 
the effect of the ionization front when interacting with density or ionization 
fluctuations occurring in the circumstellar gas (see, for instance, Soker 1998).

Symmetric LIS should be analyzed deeply, since they certainly 
gives us clues to their formation processes. In this way, a precise 
definition of the structures we are talking about is necessary. 
Hereafter we refer to as ``jets''
 the high-collimated LIS which: are not isolated knots 
but are extended in the radial direction from the central star; appear in 
opposite symmetrical pairs; and move with velocities substantially larger than 
those of the main shell, whose typical velocities are of 20 - 40 km s$^{-1}$. 
On the contrary, the features with the morphological appearance of jets, but 
which move with velocities similar to those of the main shell, or without 
available kinematical data, are called jet-like structures. It is 
clear that projection effects, which are often poorly known, play a fundamental 
role in distinguishing genuine jets and jet-like LIS. Detailed spatio-kinematic 
modeling of both the main nebulae and the LIS are therefore mandatory.

\subsection{Our new data on LIS}

We obtained high-quality narrow-band images and long-slit spectra 
for a sample 
of PNe: IC 4593 (Corradi et al. 1997); NGC 3918, K 1-2 and Wray 17-1 
(Corradi et al. 1999a); NGC 6337, He 2-186 and K 4-47 (Corradi et al. 1999b); 
IC 2553 and NGC 5882 (Corradi et al., in preparation). In analyzing these data, 
via  
spatio-kinematic modeling, our goal is to determine the 3-D geometrical and 
kinematic parameters which can, in turn, constrain the LIS formation models. 

\subsection{Jets and jet-like LIS}
Our preliminary results, in particular concerning those PNe which present 
jet or jet-like LIS, are puzzling. We follow the morphological classification 
of Schwarz et al. (1993) and the similar one of Manchado et al. (1996), and 
find that most of the PNe with LIS are not spherical: ellipticals 
($\sim 23\%$); bipolars and quadrupolars ($\sim 23\%$); point-symmetric 
($\sim 27\%$); irregulars ($\sim 20\%$): or spherical ($\sim 7\%$). 
Thus LIS seen to appear in all morphological classes, but less 
frequently in the spherical one. 

In addition, part of PNe with LIS do not have kinematic data 
published, which makes a deeper analysis impossible. Others, such as, IC 4593  
(Corradi et al. 1997),  NGC 7009 (Balick et al. 1998), K 1-2 
(Corradi et al. 1999a), He 2-429 (Guerrero et al. 1999), do have kinematic 
measurements, and their jet-like LIS have very low radial velocities. After 
deprojection for  inclination the expansion 
velocity of NCG 7009, for instance, would put it in the class of jets 
(adopting the inclination angle of Reay \& Atherton 1985), but there are 
arguments against the jet-like structure of IC 4593 being close to the plane 
of sky (Corradi et al. 1997). There are PNe which do contain jets, such as 
Hb 4 (Hajian et al. 1997), NGC 3918 (Corradi et al. 1999a), K 4-47 and He 2-186 
(Corradi et al. 1999b).

What kind of features are expected from the jet formation models? If 
jets are formed by  interacting stellar winds (ISW), in the same process 
in which the main nebulae are formed, then they should 
have ages similar to that of the main shell and lie along the symmetric 
axis (or around it, if they are precessing). The case of NGC 3918 is very 
interesting, since it does have a two-sided polar jet coeval with the highly 
axisymmetrical shell (Corradi et al. 1999a). However, the jet is much older 
than the main shell in other PNe such as NGC 6881 (Guerrero \& Manchado 1998).

Our observations for NGC 3918, He 2-186 and K 4-47, reveal that the 
collimated gas along the jets is generally increasing in velocity. 
Therefore, it is possible that the linear increase of the gas expansion  
velocity is an intrinsic characteristic of the collimation process itself. Note  
that the collimation processes in the case of K 1-2 (with a close binary 
system in the center) and NGC 3918, for instance, could not be the same. 
An increase in the expansion velocity is the expected behavior of the jets 
formed in the models of G\'arcia-Segura et al. (1999), in which the magnetic 
field of the central star is responsible for the jet collimation. 

\subsection{Other symmetrical LIS}
Here we include all the features that appear in radially symmetrical pairs 
with respect to the central star (not only along the major axis, 
but also along other directions), and those that are point-symmetrics; 
but not those clearly related to jets. One thirdy of the 
sample present this kind of symmetry: NGC 6751 (Chu et al. 1991); Cn 1-5, 
NGC 2553 (Corradi et al. 1996); NGC 3242 (Balick et al. 1998); NGC 5189, 
NGC 6826 (Phillips \& Reay 1983); NGC 6337, Wray 17-1 (Corradi et al. 1999a); 
and others.

What physical processes could be responsible for these structures? The idea 
of bullets ejected by the central star is very appealing indeed. However, 
there are problems with this model, since ballistic motions are clearly 
difficult to maintain in a hydrodynamical environment. MHD models of 
Garcia-Segura et al. (1999, and this volume) could apply here too. The latter 
could explain these symmetrical LIS only if for any reason the jet emission is  
obscured, except in the jet heads. In fact, this is not an unlikely possibility.
 New models presented in this meeting -- the stagnation 
knots from partially collimated outflows -- appear to reproduce symmetrical 
LIS, which appear not related to jets, such as the FLIERs of NGC 3242 and 
NGC 7662 (Balick et al. 1998), and in this model knots are formed in the 
zone of stagnation of bipolar shells (Steffen \& L\'opez, this volume).

\section{Conclusions}
We have briefly discussed some of the very puzzling  
characteristics of low-ionization structures in PNe. The need for 
a good determination of the basic parameters and detailed constraints to the 
LIS formation models is strongly emphasized. In addition, the predictions 
of current models, when compared to the observations, is far 
from satisfactory, and  detailed modeling of LIS is clearly a matter that
requires urgent attention. 
To date, the models for the 
jet formation which best agree with observations are those that 
consider a stellar magnetic field, even though there is no real evidence 
for the presence of magnetic fields in post-AGB stars or PNe. What kind of 
direct or even indirect evidence for these fields one would expect? Finally 
we would like to raise one more question. Since the majority of the LIS lie 
in non-spherical PNe, are the processes responsible for 
the formation of LIS the same processes that are causing the asphericity?

\acknowledgements
The work of RLMC, EV, and AM is supported by a grant of the
Spanish DGES PB97-1435-C02-01, and that of DRG by a grant from the Brazilian
Agency FAPESP (proc 98/7502-0).

\end{document}